\documentclass{article}
\usepackage{spconf,amsmath,graphicx,hyperref}
\usepackage{url}
\usepackage{bbding}
\usepackage{multirow}
\usepackage{algorithm}
\usepackage{algorithmic}
\usepackage{colortbl}
\usepackage{cite,xcolor}
\usepackage{diagbox}
\usepackage{booktabs}
\usepackage{amsfonts}
\usepackage{hhline}
\usepackage{subfig}
\usepackage{tikz}
\usetikzlibrary{arrows.meta,positioning,calc,fit,shapes.geometric}
\usepackage{CJKutf8}

\usepackage{pifont}

\newcommand{\zh}[1]{
  \begin{CJK*}{UTF8}{gbsn}#1
\end{CJK*}}

\title{TS-SP: Learning Speaker-Preserving Representations in Audio Large Language Models}
\name{Junjie Li$^1$, Zheng Liang$^1$, Zhe Li$^2$, Tianchi Liu$^3$, Kong Aik Lee$^{1,\dagger}$ \thanks{$\dagger$: Corresponding author. }}
\address{$^1$ Department of Electrical and Electronic Engineering,\\ The Hong Kong Polytechnic University, Hong Kong SAR, China \\
  $^2$ Speech, Language, and Cognition Laboratory,\\ The University of Hong Kong, Hong Kong SAR, China \\
  $^3$ National University of Singapore
}

\begin{document}
\ninept
\maketitle
\begin{abstract}
  Audio large language models (ALLMs) can understand speech content, yet their ability to use speaker identity for verification remains limited. We propose TS-SP (Two-Stage Speaker Preservation), a parameter-efficient framework for learning speaker-preserving representations and making them accessible to an ALLM's language-model component. We instantiate and evaluate TS-SP on Qwen2.5-Omni-7B. First, we adapt the audio encoder with speaker identity supervision. We then freeze the adapted encoder and train the language model to compare speakers. Both stages use low-rank adaptation (LoRA), keeping the pretrained base weights fixed. On Vox1-O, TS-SP reduces the equal error rate (EER) from 7.01\% for the Paired Loss Adaptation Baseline to 4.37\%. EER remains within 4.31--4.79\% under unseen prompts. Cross-domain evaluation on CN-Celeb yields a similar EER to the baseline, but lower accuracy at the native decision threshold. These findings support two-stage adaptation for improving speaker verification on the evaluated backbone.
\end{abstract}

\begin{keywords}
  audio large language model, speaker identity preservation, speaker representation
\end{keywords}

\section{Introduction}
\label{sec:intro}

Audio large language models (ALLMs) have shown strong performance on tasks such as speech recognition, speech translation, and audio question answering~\cite{tang2024salmonn,chu2024qwen2audio}. Models such as Qwen2.5-Omni also support spoken instruction following and natural speech generation~\cite{xu2025qwenomni}. However, these capabilities do not guarantee reliable speaker identity discrimination. Ren et al. and Thebaud et al. report weak zero-shot speaker verification performance~\cite{ren2025can,thebaud2026speaker}, while SpeakerSleuth finds that models struggle to judge speaker consistency across dialogue turns~\cite{lee2026speakersleuth}.

Multi-user voice interactions require distinguishing speakers, for example when a shared assistant decides whether to disclose private information~\cite{wang2026voxprivacy}. We study speaker identity preservation: retaining identity-discriminative audio cues and enabling the language model to use them for speaker-related decisions.

Dedicated speaker verification systems learn utterance-level speaker
embeddings through speaker-discriminative training and temporal
aggregation~\cite{snyder2017embeddings,desplanques20_interspeech,li2025xi+}.
We bring this representation-learning objective into an ALLM's native
audio encoder, without introducing an external speaker encoder.

CoLMbo conditions speaker descriptions on speaker embeddings and prompts~\cite{baali2025colmbo}, while SpeakerCard-1M provides evidence-grounded bilingual descriptions, speaker--text retrieval, and audio--description consistency tests (AC-Verify)~\cite{peng2026speakercard}. Ren et al. fine-tune Kimi-Audio through pairwise audio question answering and rule-based hard-pair sampling~\cite{ren2025can}. Other approaches supply externally pretrained speaker representations to language models (LMs): Thebaud et al. project frozen ECAPA-TDNN embeddings into the LM input space~\cite{thebaud2026speaker}, while SpeakerLLM uses a hierarchical tokenizer for utterance- and frame-level features from a frozen ReDimNet encoder~\cite{nam2026speakerllm}.

Savelyeva et al. use low-rank adaptation (LoRA)~\cite{hu2022lora} to adapt SALMONN with a joint objective combining spoof detection, three-way spoofing-aware verification, and an angular-margin speaker loss on mean-pooled QFormer features from bona fide utterances~\cite{savelyeva2026spoofing}. In contrast, TS-SP applies speaker supervision within the native audio encoder before separate pairwise LM adaptation, keeping all pretrained base weights frozen.

TS-SP first trains audio LoRA with intermediate speaker supervision, then freezes the encoder and trains LM LoRA for speaker comparison. Inference uses the native audio-to-language pathway without the speaker head or an additional connector. On Qwen2.5-Omni-7B, TS-SP improves speaker verification (SV) over the same-backbone Paired Loss Adaptation Baseline on Vox1-O/E/H. Equal error rates (EERs) remain similar across tested prompts, while cross-domain EER on CN-Celeb is comparable to the baseline.

\section{Method}
\label{sec:method}

\begin{figure*}[t]
  \centering
  \includegraphics[width=\textwidth,trim=20bp 185bp 20bp 65bp,clip]{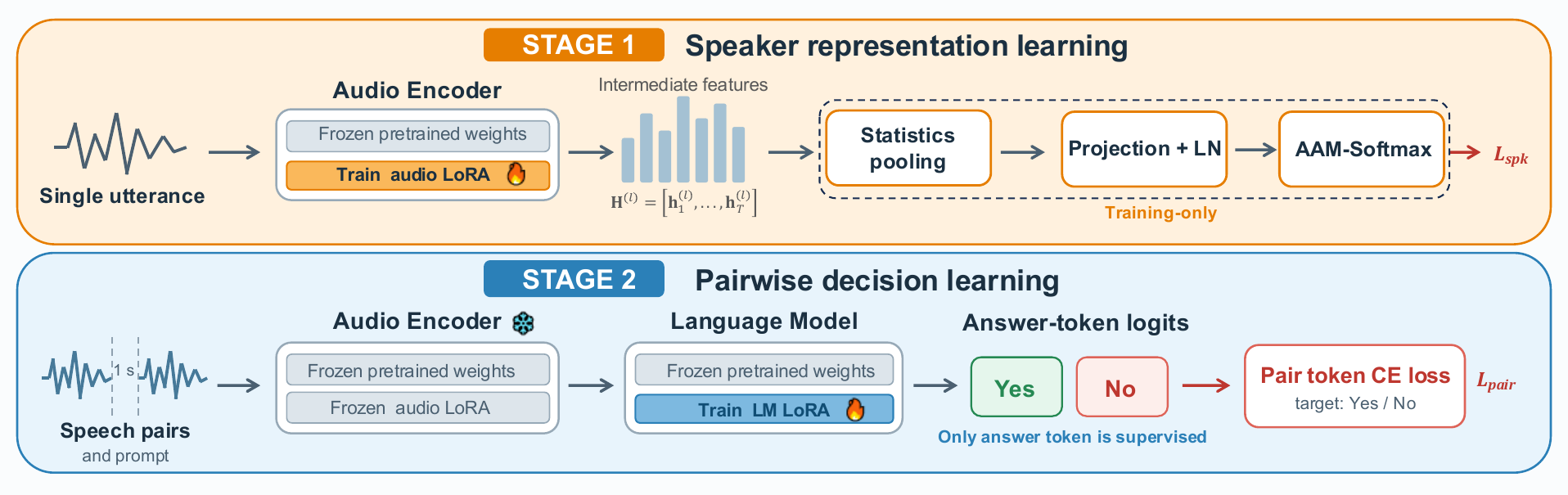}
  \caption{TS-SP: Stage~1 learns audio LoRA with intermediate speaker supervision; Stage~2 freezes the adapted audio pathway and trains LM LoRA for speaker comparison. All pretrained base weights remain fixed.}
  \label{fig:method-overview}
  \vspace{-3mm}
\end{figure*}

\subsection{Backbone and Adaptation Scope}

TS-SP requires accessible intermediate audio features and separately adaptable audio and language components. In Qwen2.5-Omni-7B, the audio encoder supplies features to the text-generating decoder in Thinker, which we call the LM~\cite{xu2025qwenomni}. Audio and LM LoRA adapters are trained in succession with all base weights fixed (Fig.~\ref{fig:method-overview}). Other backbones remain to be evaluated.

\subsection{Stage 1: Speaker-Supervised Audio Encoder}
\label{sec:training}

Let $\mathbf{H}^{(l)}=[\mathbf{h}^{(l)}_1,\ldots,\mathbf{h}^{(l)}_T]$, $\mathbf{h}^{(l)}_t\in\mathbb{R}^{d_h}$, denote the output of audio encoder layer $l$, with $T$ frames and feature dimension $d_h$. We use parameter-free statistics pooling~\cite{snyder2017embeddings}, concatenating the temporal mean and population standard deviation over valid frames. Padding is masked out; all valid frames contribute equally, without attention or learned frame weighting in the pooling operation. The trainable speaker head projects this $2d_h$-dimensional vector to $d_e$ dimensions, applies LayerNorm and $L_2$ normalization, and classifies the $C$ training speakers using additive angular-margin softmax (AAM-Softmax)~\cite{deng2019arcface}, with scale $a$ and margin $m$. For speaker label $y\in\{1,\ldots,C\}$, the loss is
\begin{equation}
  \mathcal{L}_{\mathrm{spk}}
  =-\log\frac{e^{a\cos(\theta_y+m)}}
  {e^{a\cos(\theta_y+m)}+\sum_{\substack{j=1\\j\ne y}}^{C}e^{a\cos\theta_j}},
\end{equation}
where $\theta_j$ is the angle between the normalized embedding and the normalized weight vector for class $j$. Only the audio encoder LoRA parameters up to and including layer $l$ and the speaker head are optimized by this loss; statistics pooling has no trainable parameters.

\subsection{Stage 2: Pairwise Language-Model Adaptation}
\label{sec:stage2}

Following the audio-concatenation strategy of Ren et al.~\cite{ren2025can}, we concatenate audio segments $A$ and $B$ with one second of silence between them for each verification trial. During Stage~2 training, we use the prompt: ``\textit{Are these two audio segments spoken by the same speaker? Answer only Yes or No.}'' The target is \texttt{Yes} for a same-speaker (target) trial and \texttt{No} for a different-speaker (nontarget) trial. The pair loss is
\begin{equation}
  \mathcal{L}_{\mathrm{pair}}=-\log P(t\mid A,B,q),
\end{equation}
where $q$ is the verification prompt and $t\in\{\texttt{Yes},\texttt{No}\}$ is the ground-truth answer token. Only this answer token contributes to the cross-entropy loss; prompt and audio-input positions are excluded. Stage~2 updates only the language-model LoRA parameters, keeping the adapted audio encoder frozen. The speaker head is not used in the pairwise forward path.

At inference, the answer-token probabilities provide a continuous verification score, as defined in Section~\ref{sec:metrics}.

\section{Experiments}
\label{sec:experiments}

\begin{table*}[t]
  \centering
  \caption{SV results on Vox1-O/E/H and CN-Celeb. Only the Paired Loss Adaptation Baseline and TS-SP undergo task-specific fine-tuning (VoxCeleb2); other systems are evaluated zero-shot. CN-Celeb uses 348,430 trials (1,776 target, 346,654 nontarget), without domain adaptation. Lower EER/minDCF and higher ACC are better; ACC uses $s=0$. Bold marks the best value per metric and trial list.}
  \label{tab:results}
  \setlength{\tabcolsep}{4pt}
  \resizebox{\textwidth}{!}{%
    \begin{tabular}{llccccc}
      \toprule
      System & Adaptation & EER (\%) & minDCF & Target ACC (\%) & Nontarget ACC (\%) & Overall ACC (\%)\\
      \midrule
      \rowcolor{gray!35}
      \multicolumn{7}{c}{\textbf{IN-DOMAIN EVALUATION}}\\
      \rowcolor{gray!18}
      \multicolumn{7}{l}{\textbf{Vox1-O}}\\
      Kimi-Audio-7B-Instruct~\cite{ding2025kimiaudio} & Zero-shot & 62.26 & 1.000 & 43.44 & 32.94 & 38.19\\
      Qwen2-Audio-7B-Instruct~\cite{chu2024qwen2audio} & Zero-shot & 33.39 & 1.000 & \textbf{100.00} & 0.00 & 50.00\\
      Qwen2.5-Omni-7B~\cite{xu2025qwenomni} & Zero-shot & 32.03 & 1.000 & 60.99 & 72.55 & 66.77\\
      Paired Loss Adaptation Baseline~\cite{ren2025can} & Fine-tuned & 7.01 & 0.915 & 97.99 & 87.77 & 92.88\\
      TS-SP (Ours) & Fine-tuned & \textbf{4.37} & \textbf{0.611} & 97.49 & \textbf{93.82} & \textbf{95.66}\\
      \midrule
      \rowcolor{gray!18}
      \multicolumn{7}{l}{\textbf{Vox1-E}}\\
      Kimi-Audio-7B-Instruct & Zero-shot & 62.90 & 1.000 & 42.07 & 33.10 & 37.59\\
      Qwen2-Audio-7B-Instruct & Zero-shot & 32.72 & 1.000 & \textbf{100.00} & 0.00 & 50.00\\
      Qwen2.5-Omni-7B & Zero-shot & 31.53 & 1.000 & 55.43 & 76.70 & 66.07\\
      Paired Loss Adaptation Baseline & Fine-tuned & 7.64 & 0.913 & 96.89 & 88.04 & 92.47\\
      TS-SP (Ours) & Fine-tuned & \textbf{4.41} & \textbf{0.598} & 97.22 & \textbf{93.75} & \textbf{95.48}\\
      \midrule
      \rowcolor{gray!18}
      \multicolumn{7}{l}{\textbf{Vox1-H}}\\
      Kimi-Audio-7B-Instruct & Zero-shot & 57.17 & 1.000 & 41.78 & 43.92 & 42.85\\
      Qwen2-Audio-7B-Instruct & Zero-shot & 45.98 & 1.000 & \textbf{100.00} & 0.00 & 50.01\\
      Qwen2.5-Omni-7B & Zero-shot & 44.70 & 1.000 & 55.62 & 54.96 & 55.29\\
      Paired Loss Adaptation Baseline & Fine-tuned & 19.31 & 0.999 & 97.16 & 52.77 & 74.97\\
      TS-SP (Ours) & Fine-tuned & \textbf{9.43} & \textbf{0.834} & 97.46 & \textbf{75.42} & \textbf{86.44}\\
      \midrule
      \rowcolor{gray!35}
      \multicolumn{7}{c}{\textbf{CROSS-DOMAIN EVALUATION: CN-Celeb (10\% subset)}}\\
      Qwen2.5-Omni-7B & Zero-shot & 31.93 & 1.000 & 34.01 & \textbf{90.00} & \textbf{89.72}\\
      Paired Loss Adaptation Baseline & Fine-tuned & \textbf{16.49} & 0.985 & 99.38 & 49.66 & 49.92\\
      TS-SP (Ours) & Fine-tuned & 16.74 & \textbf{0.984} & \textbf{99.61} & 34.56 & 34.89\\
      \bottomrule
    \end{tabular}
  }
  \vspace{-3mm}
\end{table*}

\subsection{Compared Systems}

The zero-shot comparison covers Kimi-Audio-7B-Instruct~\cite{ding2025kimiaudio}, Qwen2-Audio-7B-Instruct~\cite{chu2024qwen2audio}, and Qwen2.5-Omni-7B~\cite{xu2025qwenomni}. Fine-tuned systems include the Paired Loss Adaptation Baseline and TS-SP, both built on Qwen2.5-Omni-7B. Our baseline follows the pairwise audio question-answering paradigm of Ren et al.~\cite{ren2025can}, but uses our backbone, pairs, prompt, and LoRA configuration. It is not a direct reproduction; its scores in Table~\ref{tab:results} are obtained here.

\subsection{Training Details}
\label{sec:training-details}

\noindent\textbf{Training data and supervision.}
The baseline and both TS-SP stages use only the original VoxCeleb2 development recordings~\cite{chung2018voxceleb2}, without data augmentation or speed perturbation. Stage~1 uses utterances from $C=5{,}994$ speakers, with AAM-Softmax scale $a=30$ and margin $m=0.2$. In Stage~1, a forward hook extracts the output of audio encoder layer $l=30$ (zero-based indexing). These features are pooled and passed to the speaker head to compute $\mathcal{L}_{\mathrm{spk}}$. We use $d_h=1280$ and $d_e=256$, yielding a $2560$-dimensional pooled vector. The baseline and Stage~2 use the same 200,000 recording pairs: 100,000 same-speaker and 100,000 different-speaker pairs. Batch sampling maintains an approximately 1:1 target-to-nontarget ratio. Audio is resampled to 16\,kHz when necessary, and each utterance is truncated to its first 6\,s at most, without random cropping. Short inputs are right-zero-padded during feature extraction. Pairwise training uses the prompt, labels, 1\,s silence separator, and answer-only loss in Section~\ref{sec:stage2}; each concatenated pair is at most 13\,s long.

\noindent\textbf{LoRA configuration and parameter counts.}
All adaptation experiments use Qwen2.5-Omni-7B with its pretrained weights frozen. LoRA is applied to the query, key, and value projections in all 32 audio encoder layers, and to the query, key, value, and output projections in all 28 language-model layers. The audio encoder's output projections are not adapted. Both adapter sets use rank $r=16$, $\alpha=32$ (scaling factor $\alpha/r=2$), dropout 0.05, and no trainable biases. The LoRA adapters in the audio encoder and language model contain 3.932\,M and 10.093\,M parameters, respectively ($1\,\mathrm{M}=10^6$). The Paired Loss Adaptation Baseline jointly optimizes both adapter sets (14.025\,M parameters) using $\mathcal{L}_{\mathrm{pair}}$, without a speaker head or a separate speaker-pretraining stage. In TS-SP Stage~1, the speaker loss updates the adapters in audio encoder layers 0--30 (3.809\,M parameters) and the 2.191\,M-parameter speaker head, totaling 6.000\,M parameters on the loss path. The final-layer adapters lie outside this path and remain at initialization. TS-SP Stage~2 then freezes the adapted audio pathway and trains only the 10.093\,M language-model adapter parameters. Both systems retain the same number of LoRA parameters (14.025\,M) at inference, but differ in how these parameters are optimized.

\noindent\textbf{Optimization and stage transition.}
All adaptation configurations use AdamW~\cite{loshchilov2019adamw} with a learning rate of $10^{-5}$ for LoRA parameters, weight decay 0.01, and linear warmup followed by cosine decay. Training uses two GPUs with per-GPU batch size 2 and eight gradient-accumulation steps, giving an effective batch size of 32. The speaker head in TS-SP Stage~1 uses a learning rate of $10^{-4}$. TS-SP Stage~2 is initialized from the Stage~1 checkpoint after three epochs (102,375 steps), with a freshly initialized optimizer and scheduler, and runs for five epochs (31,250 steps). We report results from the final checkpoint of each run rather than a checkpoint selected by validation performance.

\subsection{Inference Details}
\label{sec:inference-details}

\noindent\textbf{Evaluation trials.}
We evaluate on the cleaned Vox1-O, Vox1-E, and Vox1-H trial lists~\cite{nagrani2020voxceleb}. For cross-domain evaluation, we randomly sample 10\% of the CN-Celeb trials~\cite{fan2019cnceleb}, preserving the original target-to-nontarget ratio, without additional fine-tuning.

\noindent\textbf{Audio inputs and scoring.}
All systems use 16\,kHz audio, retaining at most the first 6\,s of each waveform, as in training. The waveforms are concatenated with 1\,s of silence and fed as a single input. Scores use Eq.~\eqref{eq:score} and the prompt in Section~\ref{sec:stage2} unless otherwise stated.

\begin{table*}[t]
  \centering
  \caption{Prompt-robustness results of TS-SP on the cleaned Vox1-O trial
    list.  P0 is the only prompt used in Stage~2 training, while P1--P5 are unseen
    test-time prompts.  Positive and negative verbalizers correspond to
    same-speaker and different-speaker trials, respectively.  Target and
  nontarget accuracies use the native threshold $s_v=0$.}
  \label{tab:prompt-robustness}
  \setlength{\tabcolsep}{4pt}
  \resizebox{\textwidth}{!}{%
    \begin{tabular}{clp{7.1cm}cccccc}
      \toprule
      ID & Usage & Prompt supplied with the audio pair & Positive & Negative & EER (\%) & minDCF & \shortstack{Target\\ACC (\%)} & \shortstack{Nontarget\\ACC (\%)}\\
      \midrule
      P0 & Train/test & Are these two audio segments spoken by the same speaker? Answer only Yes or No. & Yes & No & 4.37 & \textbf{0.611} & 97.49 & 93.82\\
      P1 & Test only & Do these two audio segments belong to the same speaker? Respond only with Yes or No. & Yes & No & 4.33 & \textbf{0.611} & 97.55 & 93.66\\
      P2 & Test only & The two audio segments were produced by the same speaker. Is this statement true or false? Answer only True or False. & True & False & 4.39 & 0.671 & 97.15 & 94.33\\
      P3 & Test only & Compare the speakers in the two audio segments. Output 1 if they are the same speaker and 2 otherwise. & 1 & 2 & 4.79 & 0.824 & 98.69 & 90.44\\
      P4 & Test only & Are the speakers in the two audio segments the same or different? Answer only Same or Different. & Same & Different & \textbf{4.31} & 0.631 & 96.30 & \textbf{95.16}\\
      P5 & Test only & \zh{这两段音频是否由同一个说话人说出？只回答“是”或“不是”。} & \zh{是} & \zh{不是} & 4.50 & 0.643 & \textbf{99.88} & 79.88\\
      \bottomrule
    \end{tabular}
  }
\end{table*}

\subsection{Metrics}
\label{sec:metrics}
For each audio pair $(A,B)$ and verification prompt $q$, we compute a continuous score from the answer-token probabilities:
\begin{equation}
  s=\log P(\texttt{Yes}\mid A,B,q)
  -\log P(\texttt{No}\mid A,B,q).
  \label{eq:score}
\end{equation}
Let $z_{\mathrm{Yes}}$ and $z_{\mathrm{No}}$ denote the logits for these single-token answers at the same prediction position. Since the softmax normalization cancels, $s=z_{\mathrm{Yes}}-z_{\mathrm{No}}$; larger scores favor the same-speaker hypothesis~\cite{thebaud2026speaker}. This token-based score is not necessarily a calibrated speaker-hypothesis likelihood ratio.

Sweeping the score threshold yields the equal error rate (EER), where false acceptance and false rejection rates are equal, and the normalized minimum detection cost (minDCF)~\cite{sadjadi2020sre19}. We use $C_{\mathrm{miss}}=C_{\mathrm{fa}}=1$ and $P_{\mathrm{tar}}=0.01$, normalizing the detection cost by $\min\{C_{\mathrm{miss}}P_{\mathrm{tar}},C_{\mathrm{fa}}(1-P_{\mathrm{tar}})\}=0.01$. Lower values are better for both metrics.

Accuracy uses the fixed native threshold $s=0$: \texttt{Yes} if $s>0$, and \texttt{No} otherwise. Let $N_{\mathrm{tar}}$ and $N_{\mathrm{non}}$ denote the numbers of same-speaker (target) and different-speaker (nontarget) trials. Let $N_{\mathrm{TP}}$ count target trials correctly assigned \texttt{Yes}, and $N_{\mathrm{TN}}$ count nontarget trials correctly assigned \texttt{No}. The accuracies (\%) are
\begin{equation}
  \begin{aligned}
    \mathrm{ACC}_{\mathrm{tar}} &= 100\,\frac{N_{\mathrm{TP}}}{N_{\mathrm{tar}}},\\
    \mathrm{ACC}_{\mathrm{non}} &= 100\,\frac{N_{\mathrm{TN}}}{N_{\mathrm{non}}},\\
    \mathrm{ACC}_{\mathrm{overall}} &= 100\,\frac{N_{\mathrm{TP}}+N_{\mathrm{TN}}}{N_{\mathrm{tar}}+N_{\mathrm{non}}}.
  \end{aligned}
  \label{eq:accuracies}
\end{equation}
Overall ACC is thus weighted by the trial-class proportions, rather than an unweighted average of target and nontarget ACC.

\section{Results}
\label{sec:results}

\subsection{Main Speaker-Verification Results}

Table~\ref{tab:results} shows weak zero-shot speaker discrimination: the best Vox1-O EER is 32.03\% for Qwen2.5-Omni-7B, and all three systems have minDCF of 1.000. At the reported precision, Qwen2-Audio assigns all trials to \texttt{Yes} at $s=0$ across Vox1-O/E/H, while Kimi-Audio obtains overall ACC below 43\% on all three lists.

On Vox1-O, the Paired Loss Adaptation Baseline reduces EER to 7.01\%. TS-SP further lowers it to 4.37\% and minDCF from 0.915 to 0.611, while increasing overall ACC from 92.88\% to 95.66\%. Its smaller target--nontarget accuracy gap is consistent with less same-speaker bias at $s=0$. These results support TS-SP as a whole, but do not isolate intermediate speaker supervision from the effects of the training schedule and additional optimization. On Vox1-E and Vox1-H, TS-SP also improves over the Paired Loss Adaptation Baseline, reducing EER from 7.64\% to 4.41\% and from 19.31\% to 9.43\%, respectively. Overall ACC increases from 92.47\% to 95.48\% on Vox1-E and from 74.97\% to 86.44\% on Vox1-H. Although TS-SP retains a stronger same-speaker bias on Vox1-H than on Vox1-O/E, its nontarget ACC improves from 52.77\% to 75.42\% over the Paired Loss Adaptation Baseline.

On CN-Celeb, both fine-tuned systems improve EER over zero-shot Qwen2.5-Omni-7B (31.93\%) and slightly reduce minDCF from 1.000. These gains do not translate into improved overall ACC at $s=0$: ACC decreases from 89.72\% to 34.89\% for TS-SP and 49.92\% for the Paired Loss Adaptation Baseline. This list contains 99.49\% nontarget trials, so always answering \texttt{No} without processing audio would achieve 99.49\% overall ACC, exceeding all three systems. Overall ACC is therefore dominated by nontarget decisions; the zero-shot model's higher ACC should not be interpreted as stronger speaker discrimination.

Comparing the two fine-tuned systems, TS-SP has a slightly higher EER than the Paired Loss Adaptation Baseline (16.74\% versus 16.49\%) and only a marginally lower minDCF (0.984 versus 0.985). Thus, the additional gains from speaker-supervised two-stage adaptation on VoxCeleb do not carry over to CN-Celeb. This limited transfer is consistent with the broader difficulty of cross-domain generalization in conventional speaker recognition~\cite{li2026u3}.

\subsection{Robustness to Prompt Variation}
\label{sec:prompt-robustness}

We evaluate the same TS-SP checkpoint on cleaned Vox1-O using the prompts in Table~\ref{tab:prompt-robustness}. Only P0 is used in Stage~2 training; P1--P5 are unseen and require no further fine-tuning. P1 rephrases the English instruction, P2--P4 change the answer format, and P5 changes both the language and verbalizers.

For a condition with positive and negative verbalizer sequences $v^{+}$ and $v^{-}$, we generalize Eq.~\eqref{eq:score} as
\begin{equation}
  s_v=\log P(v^{+}\mid A,B,q_v)-\log P(v^{-}\mid A,B,q_v),
  \label{eq:verbalizer-score}
\end{equation}
where $q_v$ is the prompt and multi-token verbalizers use summed conditional log probabilities. The decision rule assigns the positive verbalizer if $s_v>0$ and the negative verbalizer otherwise.

EER remains within 4.31--4.79\% across P0--P5, supporting transfer beyond the training prompt for the tested variants. English rephrasing has little effect on EER, while numeric answers (P3) yield the highest EER and minDCF. The Chinese prompt P5 retains an EER close to P0 (4.50\% versus 4.37\%), but shifts target/nontarget ACC to 99.88\%/79.88\%, indicating stronger same-speaker bias. The tested prompts therefore yield similar EERs without ensuring stable minDCF or fixed-threshold decisions. Because P2--P5 change both wording and verbalizers, the effects of answer format cannot be isolated; P5 additionally changes language.

\section{Conclusion and Future Work}

TS-SP improves SV on Vox1-O/E/H through two-stage LoRA adaptation of Qwen2.5-Omni-7B, without an external speaker encoder. EER varies little across tested prompts, but fixed-threshold decisions vary. On CN-Celeb, TS-SP provides no additional EER improvement over the Paired Loss Adaptation Baseline. Future work will address cross-domain generalization and explore joint training to strengthen speaker identity modeling while preserving speech recognition, content understanding, and instruction following. Retention of these capabilities remains to be evaluated in the future.

\clearpage
\raggedbottom
\bibliographystyle{IEEEbib}
\bibliography{refs}

\end{document}